\documentclass[preprint,showpacs,preprintnumbers,amsmath,amssymb]{revtex4}
\usepackage{graphicx}
\usepackage{dcolumn}
\usepackage{bm}
\usepackage{amsmath, amsthm, amssymb}

\begin{document}

\preprint{DISTA FIN 03 2006}
\title{The art of fitting financial time series with L\'evy stable distributions}
\author{Enrico Scalas}
\affiliation{Department of Advanced Sciences and Technology, Laboratory on Complex
Systems, East Piedmont University, Via Bellini 25 g, I-15100 Alessandria,
Italy}
\email{scalas@unipmn.it}
\homepage{http://www.fracalmo.org, http://www.econophysics.org,
http://www.complexity-research.org}
\author{Kyungsik Kim}
\affiliation{Department of Physics, Pukyong National University, Pusan 608-737, Korea}
\date{\today}

\begin{abstract}
This paper illustrates a procedure for fitting financial data with $\alpha$-stable distributions.
After using all the available methods to evaluate the
distribution parameters, one can qualitatively select
the best estimate and run some goodness-of-fit tests on this estimate, in order to quantitatively assess
its quality.
It turns out that, for the two investigated data sets (MIB30 and DJIA from 2000 to present), an
$\alpha$-stable fit of log-returns is reasonably good.
\end{abstract}

\maketitle




\section{Introduction}

There are several stochastic models available to fit financial price
or index time series $\{S_i\}_{i=1}^{N}$ \cite{bertram05}. Building
on the ideas presented by Bachelier in his thesis
\cite{bachelier00,cootner64}, in a seminal paper published in 1963,
Mandelbrot proposed the L\'evy $\alpha$-stable distribution as a
suitable model for price differences, $\xi = S_{i+1} - S_{i}$, or
logarithmic returns, $\xi_{\log} = \log( S_{i+1}) - \log( S_{i})$
\cite{mandelbrot63}. In the financial literature, the debate on the
L\'evy $\alpha$-stable model focused on the infinite variance of the
distribution, leading to the introduction of subordinated models
\cite{mandelbrot67,clark73,merton90}; in the physical literature,
Mantegna used the model for the empirical analysis of historical
stock-exchange indices \cite{mantegna91}. Later, Mantegna and
Stanley proposed a ``truncated'' L\'evy distribution
\cite{mantegna94,mantegna95,koponen95}, an instance of the so-called
KoBoL (Koponen, Boyarchenko and Levendorskii) distributions
\cite{schoutens03}.

L\'evy $\alpha$-stable distributions are characterized by a
power-law decay with index $0 < \alpha \leq 2$. Fitting the tails of
an empirical distribution with a power law is not simple at all.
Weron has shown that some popular methods, such as the log-log
linear regression and Hill's estimator, give biased results and
overestimate the tail exponent for deviates taken from an
$\alpha$-stable distribution \cite{weron01}.

In this paper, a method is proposed for fitting financial log-return
time series with L\'evy $\alpha$-stable distributions. It uses the
program {\tt stable.exe} developed by Nolan \cite{nolan99} and the
Chambers-Mallow-Stuck algorithm for the generation of L\'evy
$\alpha$-stable deviates \cite{chambers76,weron96,mcculloch}. The
datasets are: the daily adjusted close for the DJIA index taken from
{\tt http://finance.yahoo.com} and the daily adjusted close for the
MIB30 index taken from  {\tt http://it.finance.yahoo.com} both for
the period 1 January 2000 - 3 August 2006. The two datasets and the
program {\tt stable.exe} are freely available, so that whoever can
reproduce the results reported below and use the method on other
datasets.

The definition of L\'evy $\alpha$-stable distributions is presented
in Section II. Section III is devoted to the results of the
empirical analysis. A critical discussion of these results can be
found in Section IV.

\section{Theory}

A random variable $\Xi$ is {\em stable} or {\em stable in the broad sense} if, given two
independent copies of $\Xi$, $\Xi_1$ and $\Xi_2$, and any positive constant $a$ and $b$,
there exist some positive $c$ and some real $d$ such that the sum $a \Xi_1 + b \Xi_2$
has the same distribution as $c \Xi + d$. If this property holds with $d=0$ for any
$a$ and $b$ then $\Xi$ is called {\em strictly stable} or {\em stable in the narrow sense}.
This definition is equivalent to the following one which relates the stable property to
convolutions of distributions and to the generalization of the central limit theorem
\cite{levy}: A random variable $\Xi$ is stable if and only if, for all $n>1$
independent and identical copies of $\Xi$, $\Xi_1, \ldots, \Xi_n$, there exist a positive
constant $c_n$ and a real constant $d_n$ such that the sum $\Xi_1 + \ldots + \Xi_n$ has
the same distribution as $c_n \Xi + d_n$. It turns out that a random variable $\Xi$ is
stable if and only if it has the same distribution as $a Z + b$, where
$0< \alpha \leq 2$, $-1 \leq \beta \leq 1$, $a>0$, $b \in \mathbb{R}$ and $Z$
is a random variable with the following characteristic function:
\begin{equation}
\label{cf1}
\langle \exp(i \kappa Z) \rangle = \exp\left(
-|\kappa|^{\alpha} \left[1 - i \beta \tan \left( \frac{\pi \alpha}{2} \right) {\rm sign}
(\kappa) \right] \right) \,\, {\rm if} \, \alpha \neq 1
\end{equation}
and
\begin{equation}
\label{cf2}
\langle \exp(i \kappa Z) \rangle = \exp\left(
-|\kappa| \left[1 + i \beta \left( \frac{2}{\pi} \right) {\rm sign}
(\kappa) \log(|\kappa|) \right] \right) \,\, {\rm if} \, \alpha = 1.
\end{equation}
Thus, four parameters ($a$, $b$, $\alpha$, $\beta$) are needed to specify a general stable
distribution. Unfortunately, the parameterization is not unique and this has caused several
errors \cite{nolan98}. In this paper, a parameterization is used, due to Samorodnitsky and Taqqu
\cite{taqqu94}:
\begin{equation}
\label{cf1st}
\langle \exp(i \kappa \Xi) \rangle = \exp\left(
-\gamma^{\alpha} |\kappa|^{\alpha} \left[1 - i \beta \tan \left( \frac{\pi \alpha}{2} \right) {\rm sign}
(\kappa) \right] + i \delta \kappa \right) \,\, {\rm if} \, \alpha \neq 1
\end{equation}
and
\begin{equation}
\label{cf2st}
\langle \exp(i \kappa \Xi) \rangle = \exp\left(
-\gamma |\kappa| \left[1 + i \beta \left( \frac{2}{\pi} \right) {\rm sign}
(\kappa) \log(|\kappa|) \right] +i \delta \kappa \right) \,\, {\rm if} \, \alpha = 1.
\end{equation}
This parameterization is called $S1$ in {\tt stable.exe}. The program uses a different parameterization
(called $S0$) for numerical calculations:
\begin{equation}
\label{cf1ZM}
\langle \exp(i \kappa \Xi_0) \rangle = \exp\left(
-\gamma^{\alpha} |\kappa|^{\alpha} \left[1 - i \beta \tan \left( \frac{\pi \alpha}{2} \right) {\rm sign}
(\kappa) \right] + i \left[ \delta_0 - \beta  \tan \left( \frac{\pi \alpha}{2} \right) \right] \kappa \right) \,\, {\rm if} \, \alpha \neq 1
\end{equation}
and
\begin{equation}
\label{cf2ZM}
\langle \exp(i \kappa \Xi_0) \rangle = \exp\left(
-\gamma |\kappa| \left[1 + i \beta \left( \frac{2}{\pi} \right) {\rm sign}
(\kappa) \log(|\kappa|) \right] +i \left[ \delta_0 - \beta
\left( \frac{2}{\pi} \right) \gamma \log \gamma \right]
\kappa \right) \,\, {\rm if} \, \alpha = 1.
\end{equation}
The above equations are modified versions of Zolotarev's
parameterizations \cite{zolotarev}. Notice that in Eqs.
(\ref{cf1st}) and (\ref{cf2st}), the scale $\gamma$ is positive and
the location parameter $\delta$ has values in $\mathbb{R}$.

\section{Results}

Daily values for the MIB30 and DJIA adjusted close have been
downloaded from {\tt http://it.finance.yahoo.com} and {\tt http://finance.yahoo.com}, respectively,
for the period 1 January 2000 - 3 August 2006 \cite{adjusted}.
The MIB30 is an index comprising 30 ``blue chip'' shares on the
Italian Stock Exchange, whereas
the DJIA is a weighted average of the prices of 30 industrial companies that are representative of the
market as a whole. Notice that the MIB30 includes non-industrial shares. Therefore, the two indices cannot be
used for comparisons on the behaviour of the industrial sector in Italy and in the USA.
Moreover, the composition of indices
varies with time, making it problematic to compare different historical
periods. However, the following analysis concerns the
statistical properties of the two indices, considered representative of the stock-exchange
average trends. The datasets are presented in Figs 1-4. Figs. 1 and 2 report the
index value as a function of trading day. There are 1709 points in the MIB30 dataset and 1656 points in the
DJIA dataset. Correpondingly, there are 1708 MIB30 log-returns and 1655 DJIA log-returns. They are represented
in Figs. 3 and 4, respectively. The intermittent behaviour typical of log-return time series can
be already detected by eye inspection of Figs. 3 and 4, but this property will not be further studied.

\begin{figure}
\includegraphics{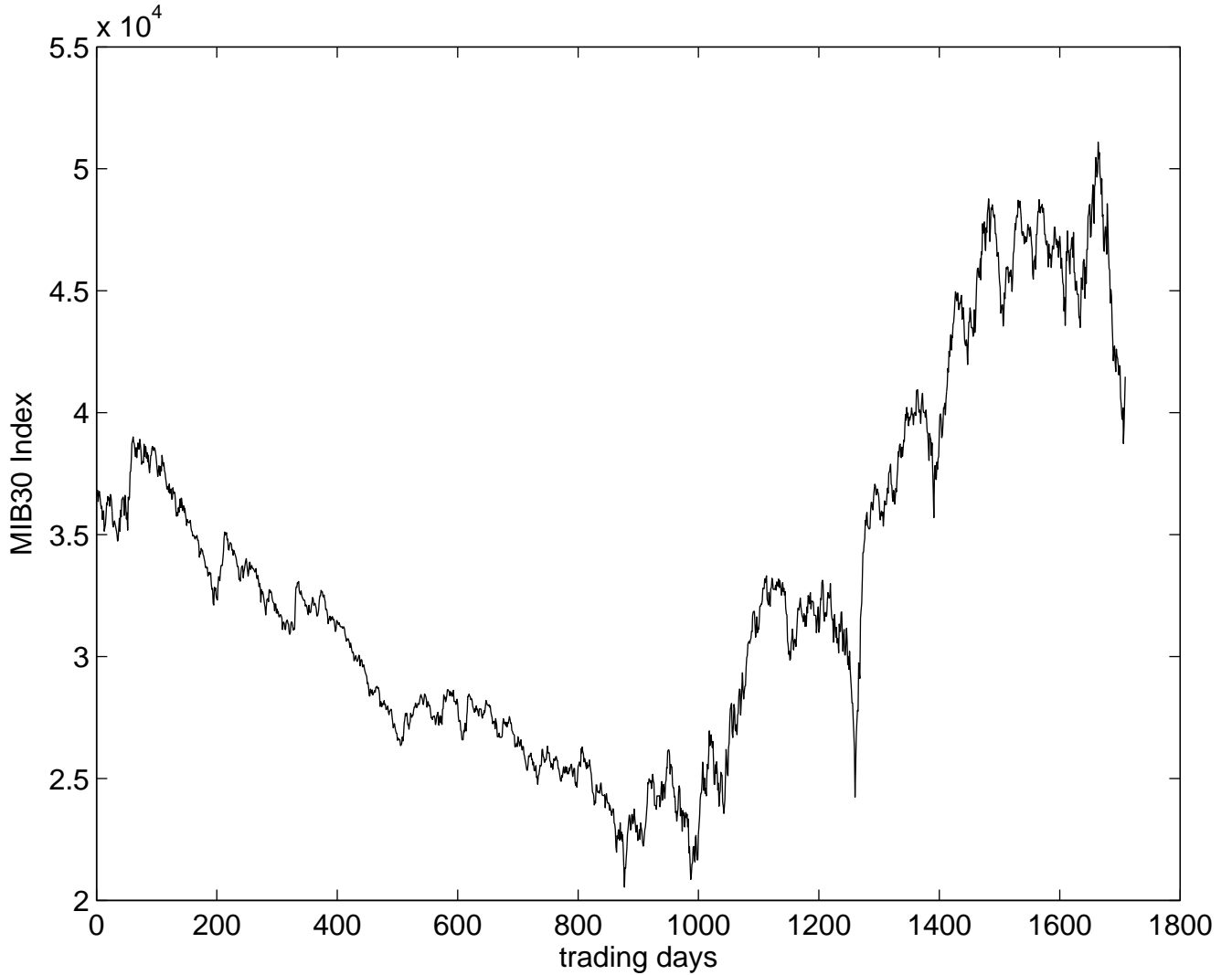}%
\caption{MIB30: Adjusted close time series. \label{Fig.1a}}
\end{figure}

\begin{figure}
\includegraphics{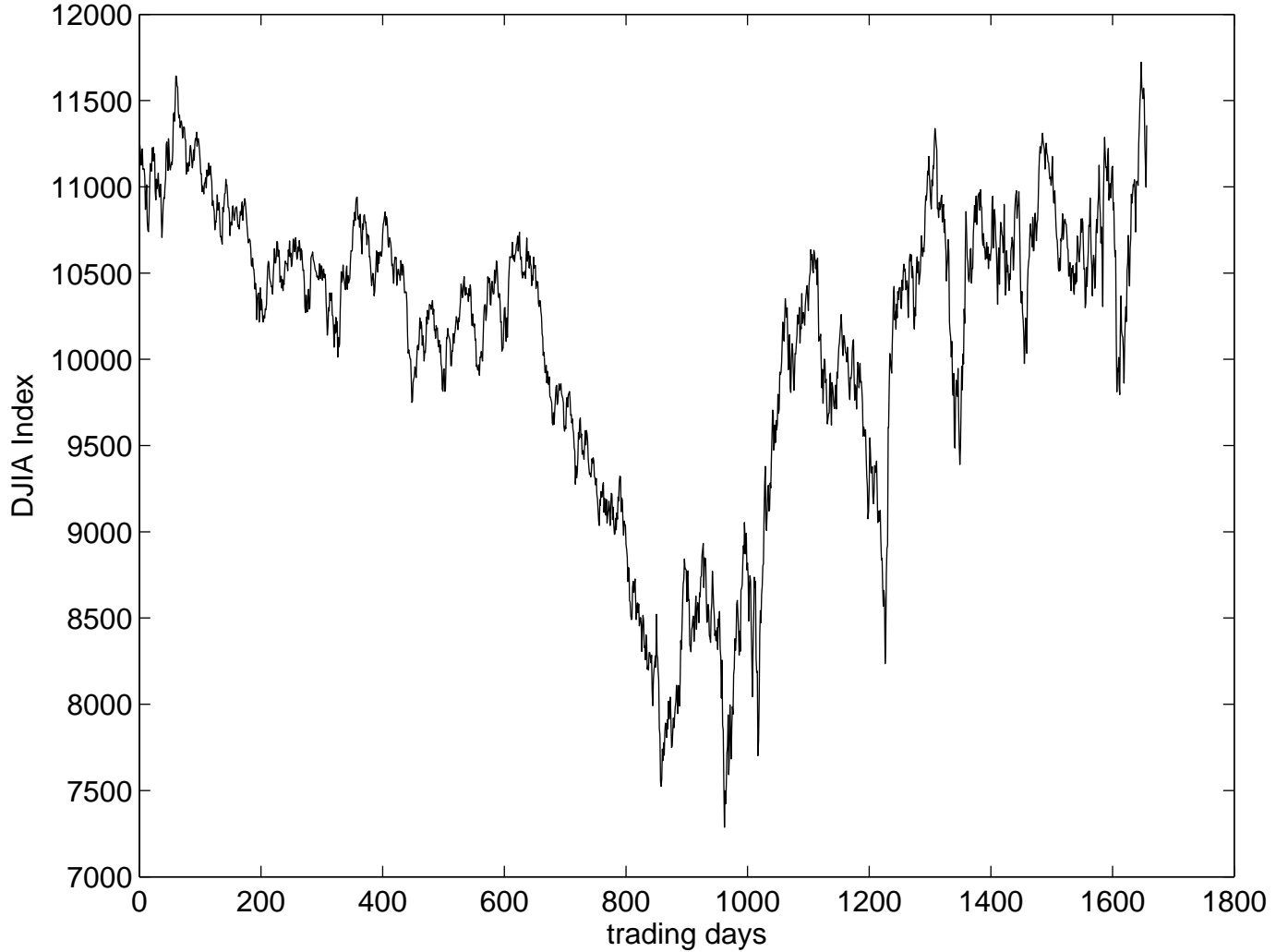}%
\caption{DJIA: Adjusted close time series.  \label{Fig.1b}}
\end{figure}

\begin{figure}
\includegraphics{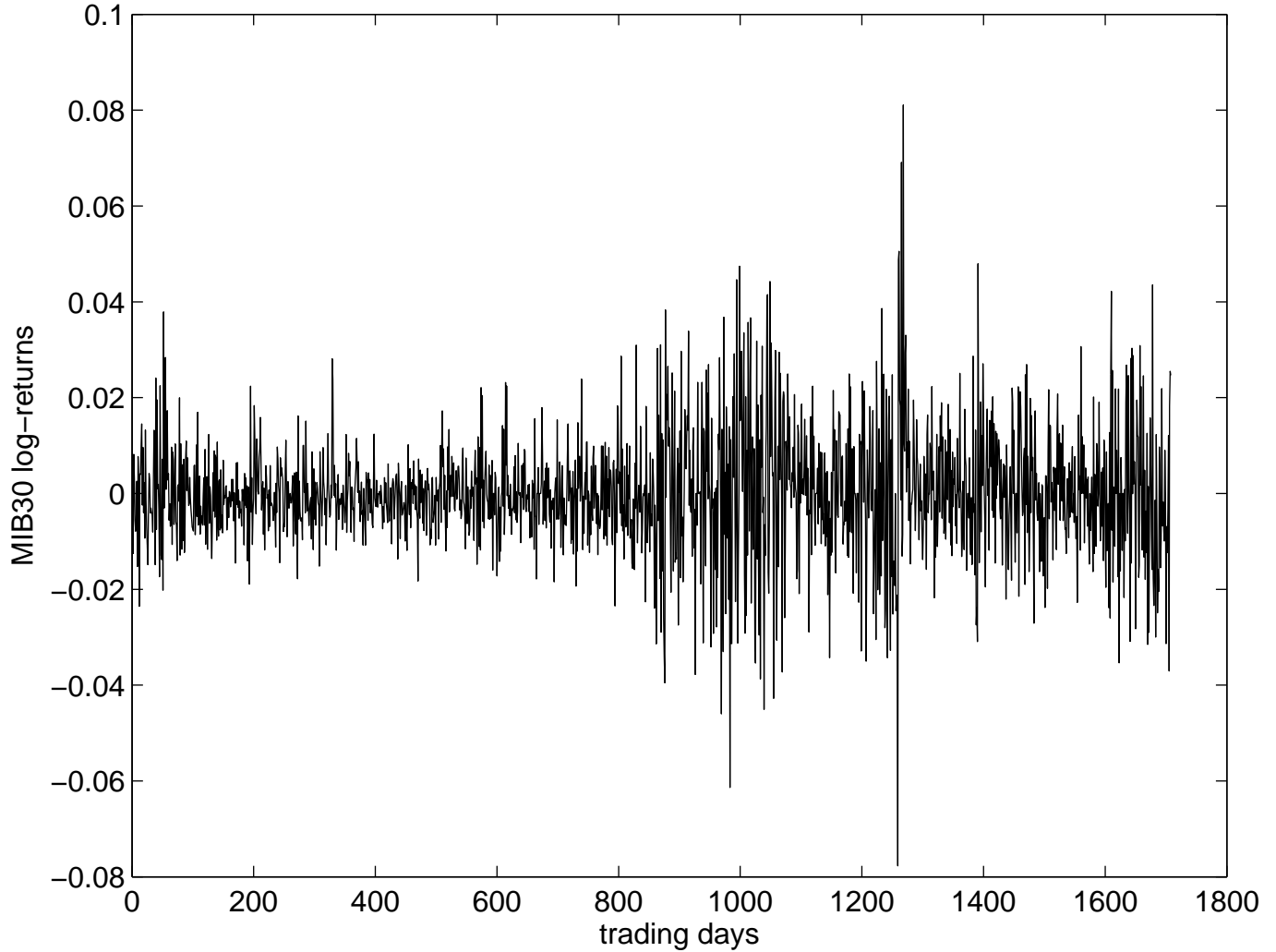}%
\caption{MIB30: Log-return time series. \label{Fig2.a}}
\end{figure}

\begin{figure}
\includegraphics{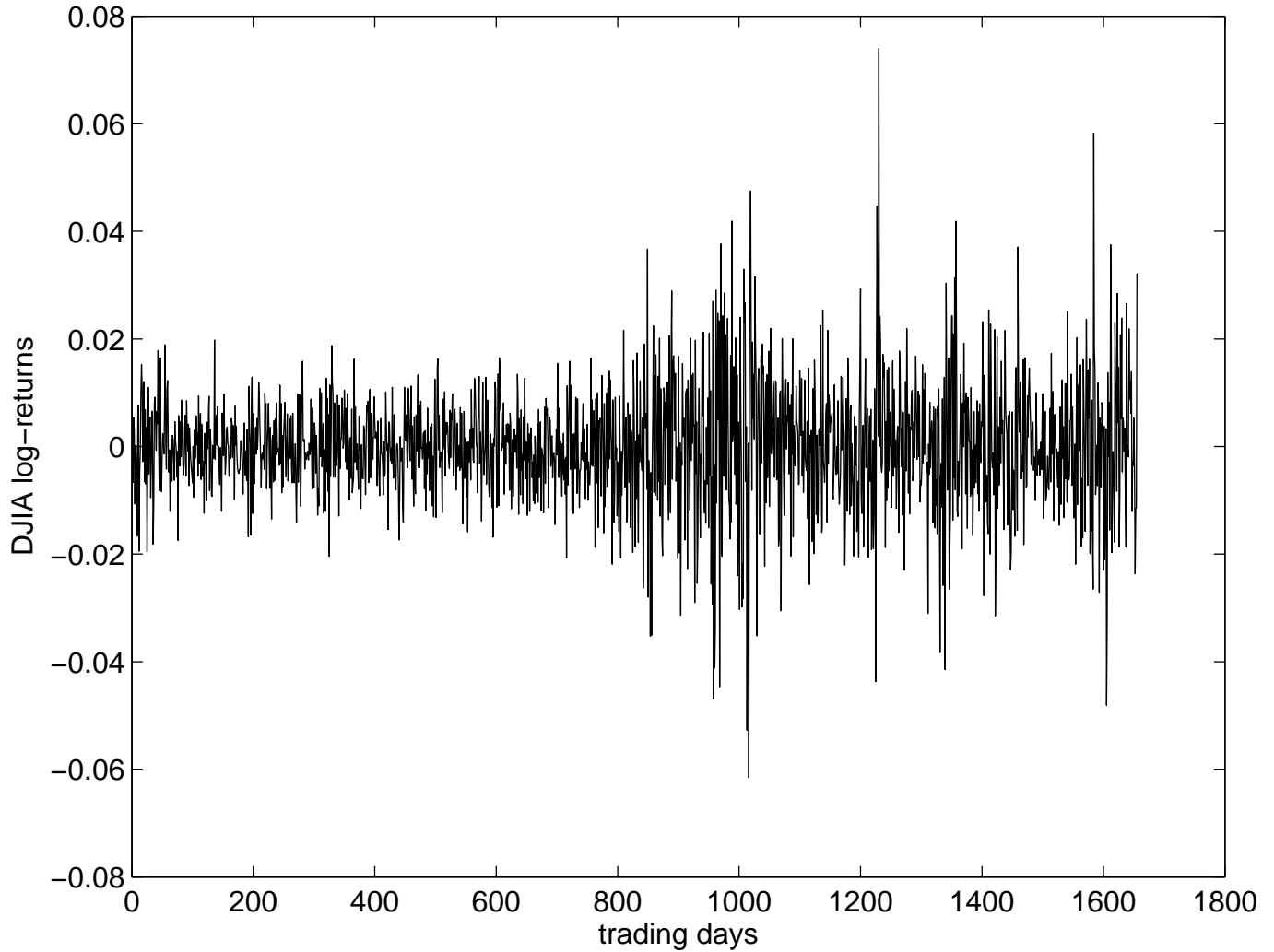}%
\caption{DJIA: Log-return time series. \label{Fig2.b}}
\end{figure}

In Table I, the mean, variance, skewness, and kurtosis are reported for both the MIB30 and the DJIA
log-return time series.
\begin{table}
\caption{\label{tab:table1} Mean, variance, skewness, and kurtosis
of the two log-return time series.}
\begin{ruledtabular}
\begin{tabular}{lcccc}
Index&Mean&Variance&Skewness&Kurtosis\\
\hline
MIB30 & $7.3 \cdot 10^{-5}$ & $1.6 \cdot 10^{-4}$ &  0.22 & 6.7\\
DJIA & $6.1 \cdot 10^{-6}$ & $1.3 \cdot 10^{-4}$ & 0.038 & 6.6 \\
\end{tabular}
\end{ruledtabular}
\end{table}

The two log-return time series were given as input to {\tt stable.exe}. The program implements
three methods for the estimate of the four parameters of Eqs. (\ref{cf1st}) and (\ref{cf2st}).
The first one is based on a maximum likelihood (ML) estimator \cite{nolan99,dumouchel83}. The second method
uses tabulated quantiles (QB) of L\'evy $\alpha$-stable distributions \cite{mcculloch86}
and it is restricted to $\alpha \geq 0.6$. Finally, in the third method, a regression on the
sample characteristic (SC) function is used \cite{koutrouvelis,kogon}.

In Table II, the estimated values of $\alpha$, $\beta$, $\gamma$, and $\delta$ are reported. The
estimates were obtained with the standard settings of {\tt stable.exe}.
\begin{table}
\caption{\label{tab:table2} Estimated parameters of the
$\alpha$-stable distribution.}
\begin{ruledtabular}
\begin{tabular}{lcccc}
Method&$\alpha$&$\beta$&$\gamma$&$\delta$\\
\hline
MIB30 & & & & \\
\hline
ML & 1.57 & 0.159 & $6.76 \cdot 10^{-3}$ & $3.50 \cdot 10^{-4}$ \\
QB & 1.42 & 0.108 & $6.23 \cdot 10^{-3}$ & $5.43 \cdot 10^{-4}$ \\
SC & 1.72 & 0.263 & $7.06 \cdot 10^{-3}$ & $2.53 \cdot 10^{-4}$ \\
\hline
DJIA & & & & \\
\hline
ML & 1.73 & 0.014 & $6.60 \cdot 10^{-3}$ & $4.43 \cdot 10^{-5}$ \\
QB & 1.60 & -0.004 & $6.21 \cdot 10^{-3}$ & $2.69 \cdot 10^{-4}$ \\
SC & 1.81 & 0.129 & $6.69 \cdot 10^{-3}$ & $1.79 \cdot 10^{-4}$ \\
\end{tabular}
\end{ruledtabular}
\end{table}
In order to preliminary assess the quality of the fits, three
synthetic series of log-returns for each index were generated with
the Chambers-Mallow-Stuck algorithm. The empirical complementary
cumulative distribution function (CCDF) for absolute log-returns was
compared with the simulated CCDFs, see Figs. 5 and 6. In both cases,
the fit based on the SC function turned out to be the best one. A
refinement of the ML method gave the same values for the four
parameters as the SC algorithm. Therefore, the SC result was
selected as the null hypothesis for two standard quantitative
goodness-of-fit tests: The one-sided Kolmogorov-Smirnov (KS) test
and the $\chi^{2}$ test.

\begin{figure}
\includegraphics{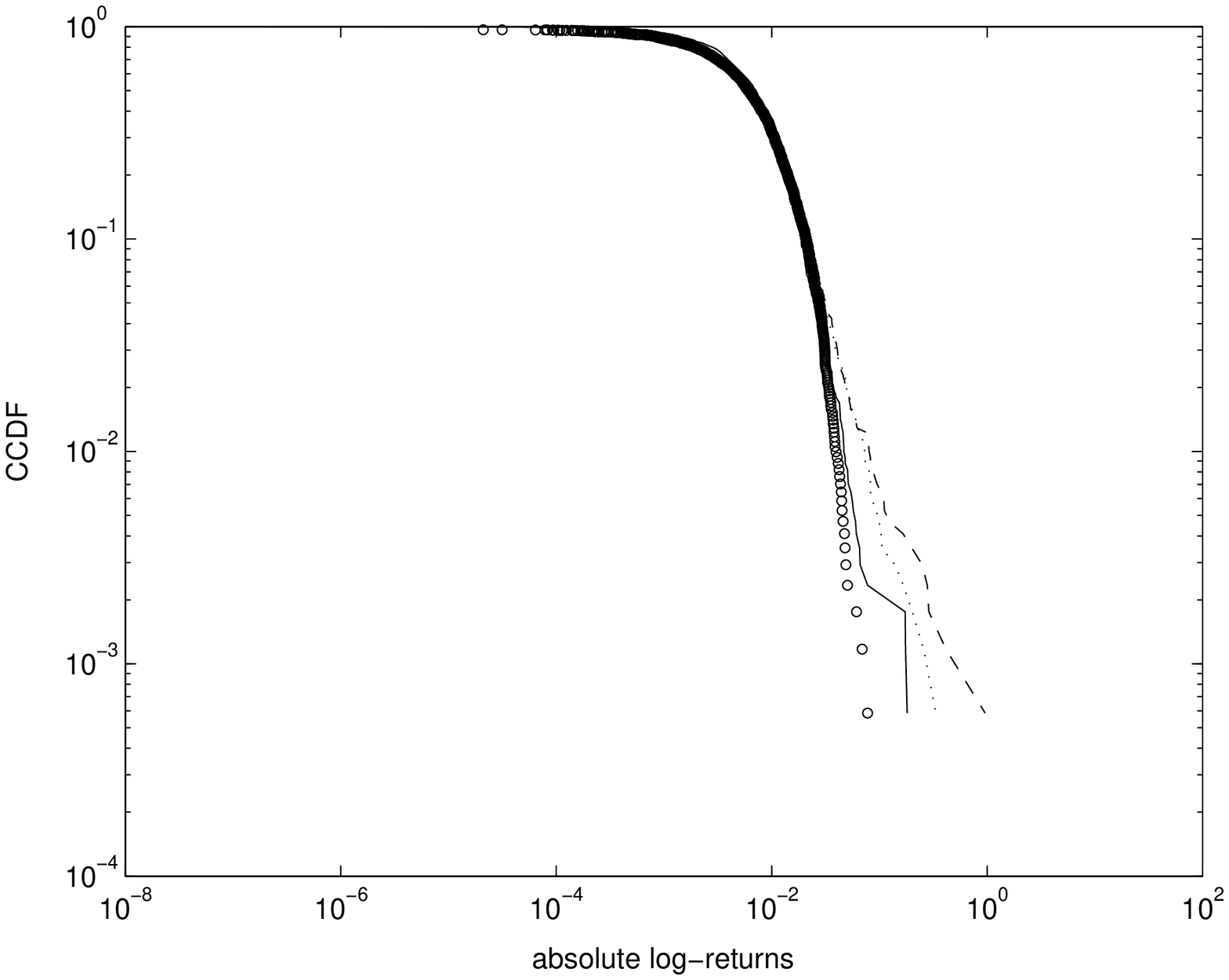}%
\caption{MIB30: Comparison of CCDFs for absolute log-returns.
Circles: Empirical data. Dotted line: Maximum-likelihood fit. Dashed
line: Quantile based fit. Solid line: Fit based on the SC function.
\label{Fig3.a}}
\end{figure}

\begin{figure}
\includegraphics{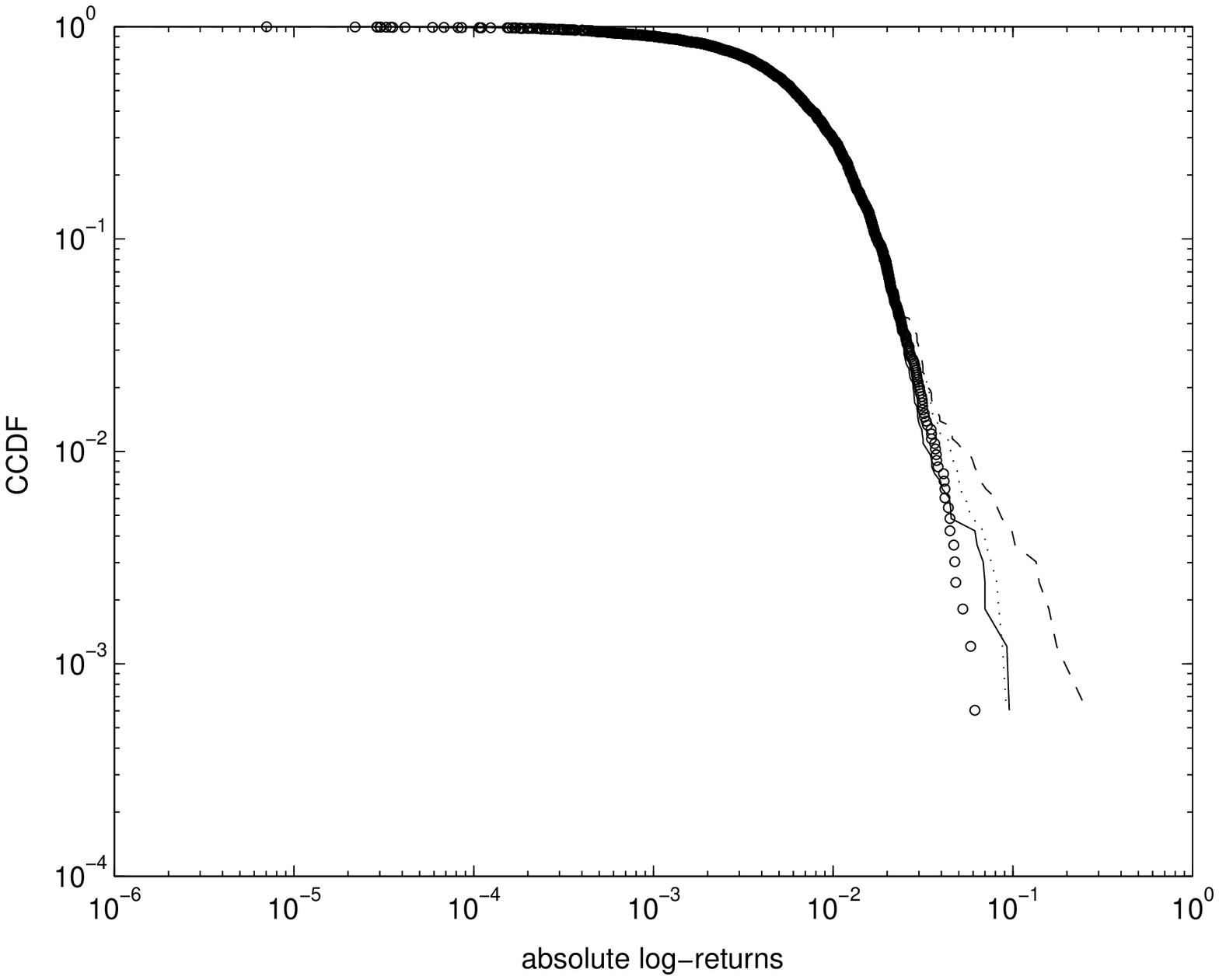}%
\caption{DJIA: Comparison of CCDFs for absolute log-returns.
Circles: Empirical data. Dotted line: Maximum-likelihood fit. Dashed
line: Quantile based fit. Solid line: Fit based on the SC function.
\label{Fig3.b}}
\end{figure}

For the KS test, the range of MIB30 log-returns,
$(-0.0777,\,0.0811)$, was divided into 1654 intervals of width $9.6
\times 10^{-5}$. Then the number of points lying in each interval
was counted and partially summed starting from $-0.0777$, leading to
an estimate of the empirical cumulative distribution function (CDF).
The same procedure was used for DJIA log-returns. In this case the
range was $(-0.0615,\,0.0740)$, the number of intervals 1693, and
their width $8.0 \times 10^{-5}$. In Figs. $7$ and $8$, the
empirical CDF is plotted together with the theoretical CDF obtained
from the fit based on the SC function. For large sample sizes, the
one-sided KS parameter, $D$, is approximately given by:
$$
D = \max(|{\rm CDF}_i - {\rm CDFth}_i|),
$$
where ${\rm CDF}_i$ and ${\rm CDFth}_i$ are, respectively, the
empirical and the theoretical values corresponding to the $i-$th
bin; at $5\%$ significance, $D$ can be compared with the limiting
value $d=1.36/\sqrt{N}$, where $N$ is the number of empirical CDF
points. For MIB30 log-returns $D=0.0387$ and $d=0.0334$, whereas for
DJIA log-returns $D=0.0232$ and $d=0.0330$. Therefore, the null
hypothesis of $\alpha$-stable distributed log-returns can be
rejected for the MIB30, but not for the DJIA data.

\begin{figure}
\includegraphics{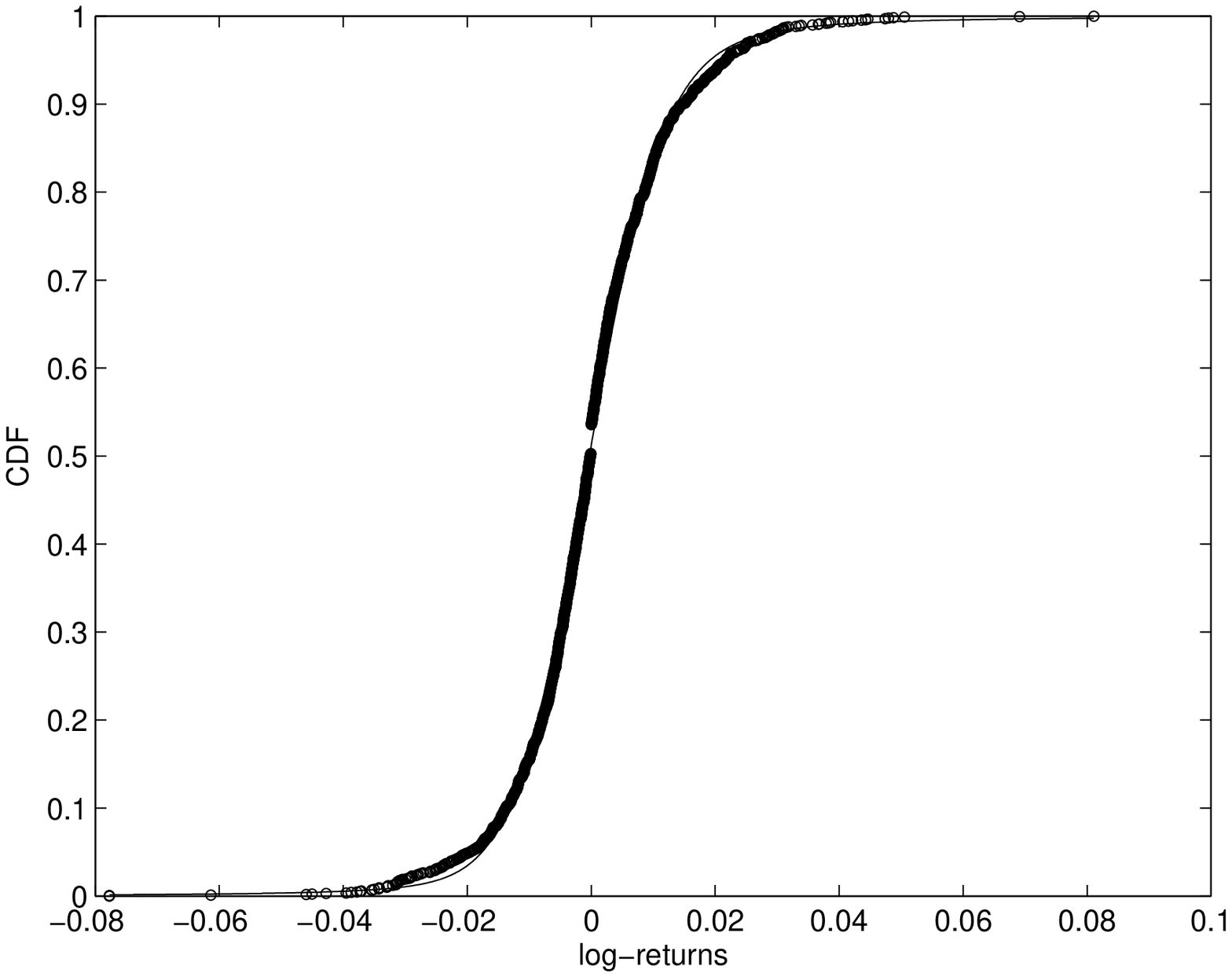}%
\caption{MIB30: Comparison of CDFs for log-returns. Circles:
Empirical data. Solid line: Fit based on the SC function.
\label{Fig4.a}}
\end{figure}

\begin{figure}
\includegraphics{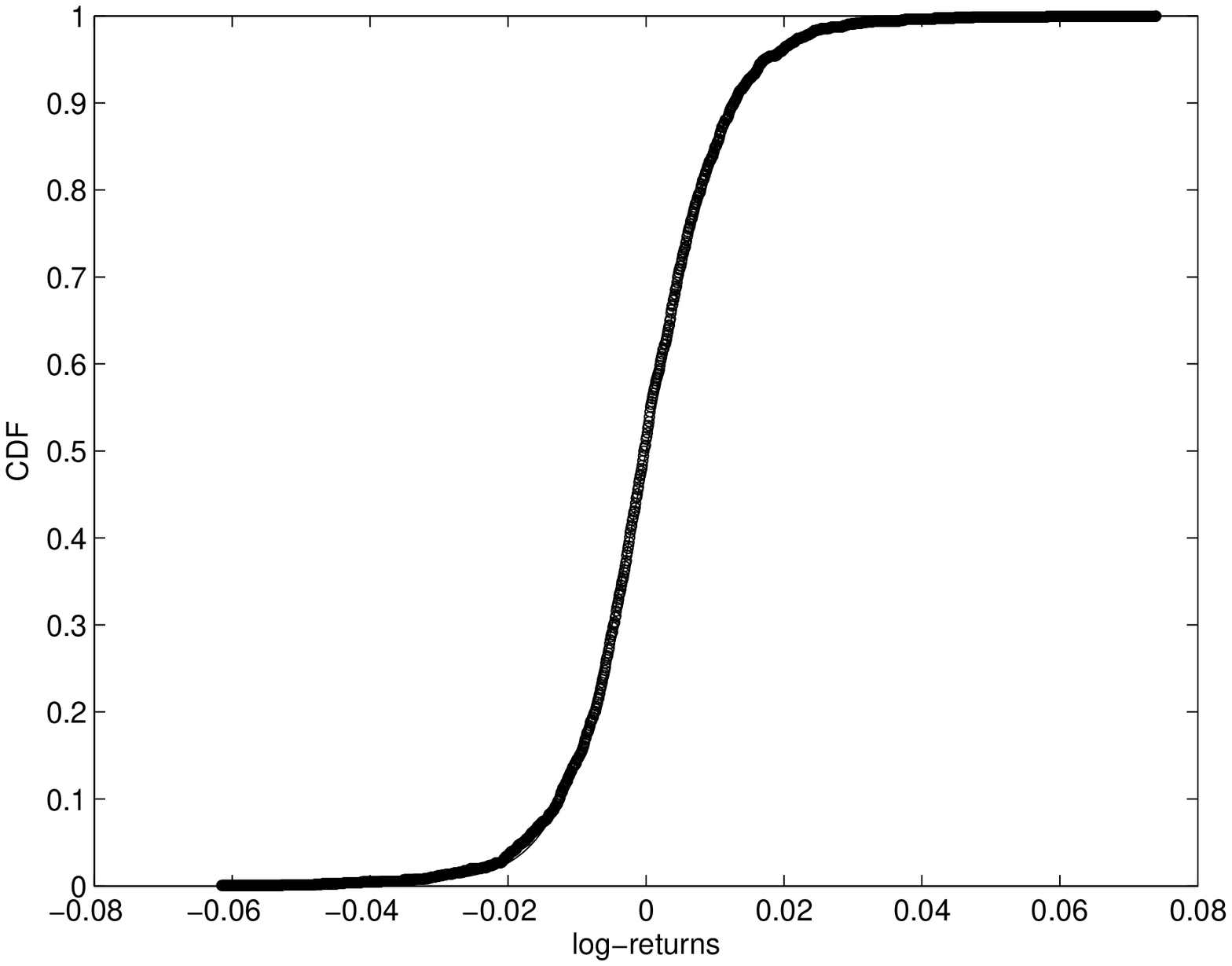}%
\caption{DJIA: Comparison of CDFs for log-returns. Circles:
Empirical data. Solid line: Fit based on the SC function.
\label{Fig4.b}}
\end{figure}

For the $\chi^{2}$ test, the range of MIB30 and DJIA log-returns was divided into 30
equal intervals. Then, the observed $O_i$ and expected $E_i$ number of points lying in each
interval were evaluated; these data are plotted in Figs. 9 and 10.
After aggregating the intervals with $E_i < 5$, $\bar{\chi}^{2}$ was obtained from the formula:
$$
\bar{\chi}^2 = \sum_i \frac{(O_i - E_i)^2}{E_i}
$$
The number of degrees of freedom is given by the number of intervals where $E_i \geq 5$ minus 5 (4 estimated
parameters and the normalization).
For MIB30 data, $\bar{\chi}^2 = 91.5$ with 10 degrees of freedom. The probability that $\chi^2 > \bar{\chi}^2$
is 0. For the DJIA time series, $\bar{\chi}^2 = 26.8$ with 11 degrees of freedom. The probability that
$\chi^2 > \bar{\chi}^2$ is 0.005. Again,
for DJIA data, even if at a low significance level, the null hypothesis may be accepted.

\begin{figure}
\includegraphics{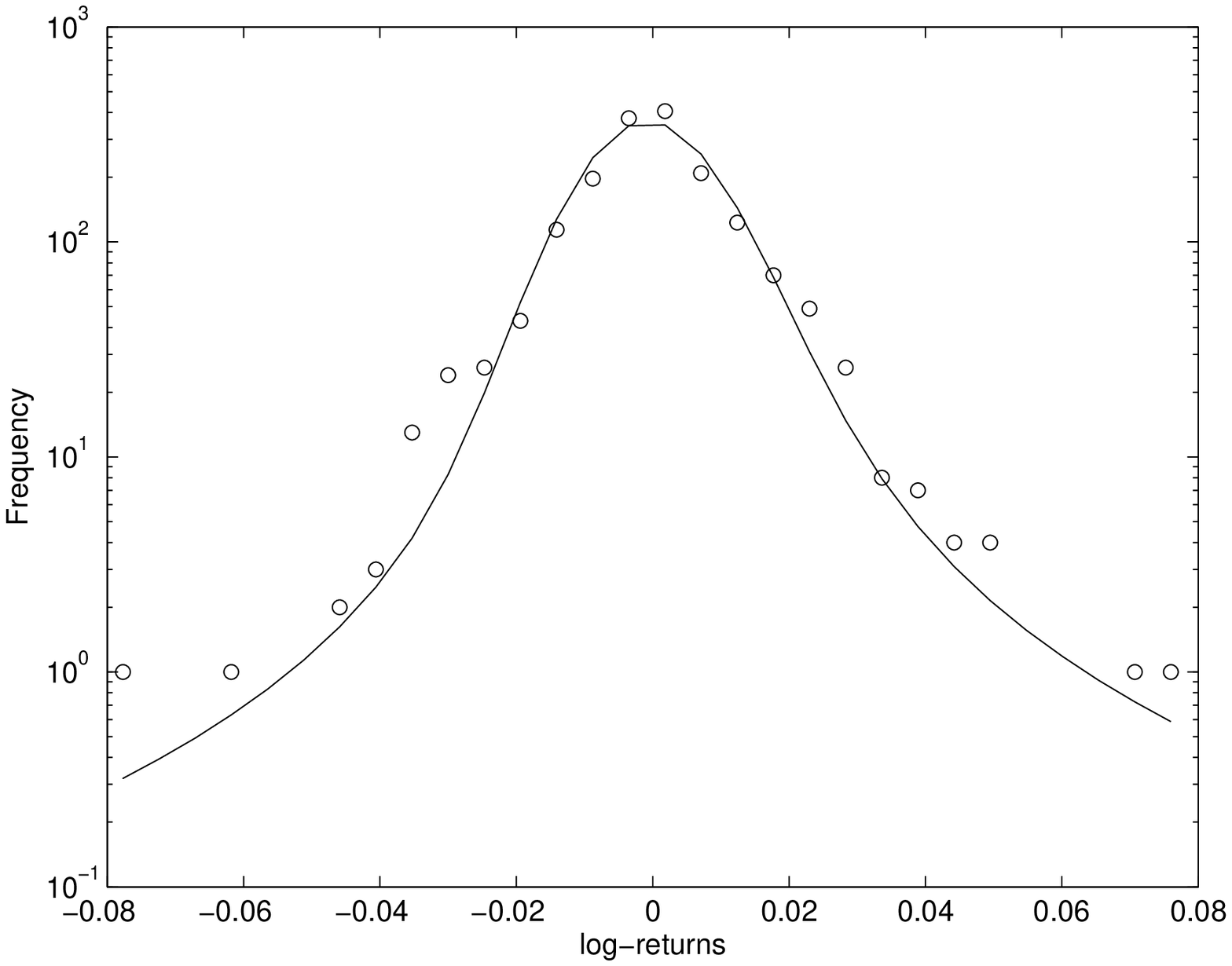}%
\caption{MIB30: Comparison of PDFs for log-returns. Circles:
Empirical data. Solid-line: Fit based on the SC function.
\label{Fig5.a}}
\end{figure}

\begin{figure}
\includegraphics{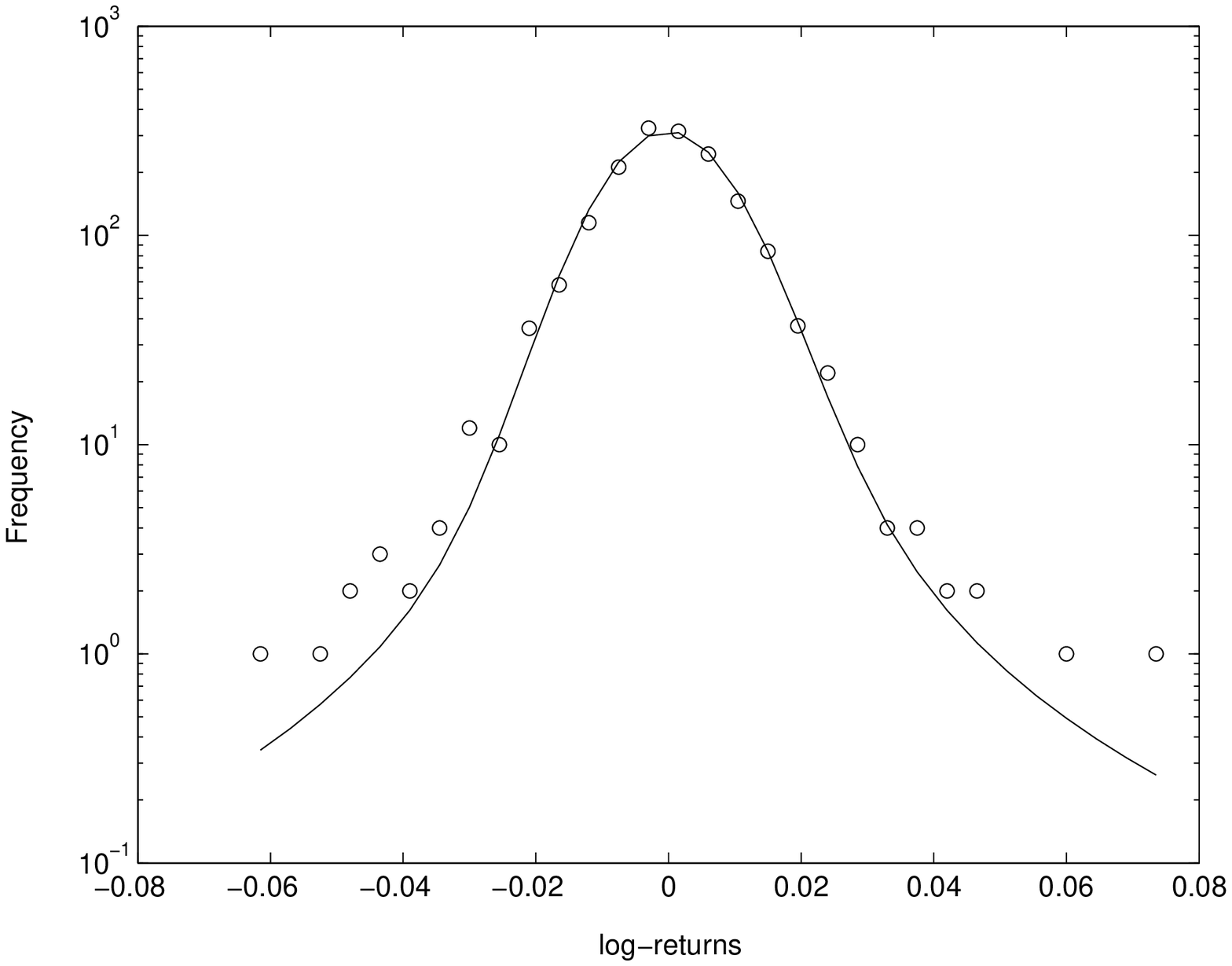}%
\caption{DJIA: Comparison of PDFs for log-returns. Circles:
Empirical data. Solid-line: Fit based on SC function.
\label{Fig5.b}}
\end{figure}

\section{Conclusions and outlook}

This paper illustrates a procedure for fitting financial data with
$\alpha$-stable distributions. The first step is to use all the
available methods to evaluate the four parameters $\alpha$, $\beta$,
$\gamma$, and $\delta$. Then, one can qualitatively select the best
estimate and run some goodness-of-fit tests, in order to
quantitatively assess the quality of the fit.

The main conclusion of this paper is that, for the investigated data
sets, an $\alpha$-stable fit is not so bad; the best parameter
estimate is obtained with a method based on a sample characteristic
function fit. Incidentally, the tail index, $\alpha$, is 1.72 for
MIB30 and 1.81 for DJIA. These values are consistent with previous
results \cite{rachev00} and with remarks made by Weron
\cite{weron01}. The performance in two standard goodness-of-fit
tests (KS and $\chi^2$) is better for DJIA data.

The two hypothesis tests used in this paper have some limitations. For instance, the KS
test is more sensitive to the central part of the distribution and underestimates the
tail contribution. For this reason, it would be better to use the Anderson-Darling (AD) test
\cite{stephens74}. However, a standardized AD test is not available for $\alpha$-stable
distributions. Moreover, the value of $\bar{\chi}^2$ in the $\chi^2$ test is sensitive to the
choice of intervals, and a detailed analysis on this dependence would be necessary.

Given a set of financial log-returns, is it possible to find the
best-fitting distribution? In general this question is ill-posed. As
mentioned in the introduction, there are several possible competing
distributions that can give very similar results in the interval of
interest. Moreover, depending on the specific criterion chosen, different
distributions may turn out to be the {\em best} according to that
criterion. Therefore, if there is no theory suggesting the choice of
a specific distribution, it is advisable to use a pragmatic and
heuristic approach, application-oriented. For example, Figs. 9 and
10 show that the L\'evy $\alpha$-stable fit discussed in this paper
tends to underestimate the tails of the probability density function
(PDF) in the two investigated cases. In risk asessment procedures,
such as value at risk estimates, this may be an undesirable feature,
and it could be wiser to look for some other probability density
whose the PDF prudentially overestimates the tail region.

\section*{ACKNOWLEDGEMENTS}

This work has been partially supported by the Italian MIUR project "Dinamica di altissima
frequenza nei mercati finanziari".

\end{document}